\documentclass{article}

\usepackage[english]{babel}

\usepackage[letterpaper,top=2cm,bottom=2cm,left=3cm,right=3cm,marginparwidth=1.75cm]{geometry}

\usepackage{amsmath}
\usepackage{graphicx}
\usepackage[colorlinks=true, allcolors=blue]{hyperref}
\usepackage{tabularx}
\usepackage{booktabs}
\usepackage{array}      
\usepackage{makecell}   
\usepackage{geometry}   
\usepackage{ragged2e}   
\usepackage{xurl}
\title{A Topological Soliton Model for Ball Lightning: Theory and 3D Gross-Pitaevskii Equation Verification}
\author{Zhe Li}

\begin{document}
\maketitle

\begin{abstract}
Ball lightning remains one of the most enigmatic atmospheric phenomena, characterized by its long lifetime, ability to penetrate materials, and stable spherical structure. Here we propose a novel theoretical framework interpreting ball lightning as a three-dimensional projection of a high-dimensional topological soliton. The system is described by a nonlinear Schrödinger equation with attractive interactions, stabilized by a non-zero topological charge. Through comprehensive numerical simulations of the three-dimensional Gross-Pitaevskii equation, we verify the model's core predictions: (1) long-lived stability protected by topological invariants, (2) low transmission probability due to wavefunction orthogonality, and (3) energy and size scales consistent with observational data. The soliton lifetime $\tau \sim \hbar/\Gamma$ naturally explains the observed second-scale durations. Our work provides a self-consistent physical explanation for ball lightning while offering concrete pathways for experimental realization of three-dimensional topological solitons in Bose-Einstein condensates and nonlinear optical systems. This theoretical framework gains additional support from recent experimental breakthroughs in laboratory generation of ball-lightning-like structures.

\end{abstract}
\section{Introduction}
Ball lightning, a rare atmospheric phenomenon manifesting as luminous spheres 10-100 cm in diameter with lifetimes up to several seconds, has perplexed scientists for centuries \cite{Stenhoff1999}. Its ability to penetrate solid materials like glass and its occasional explosive disappearance further complicate its physical understanding. Despite thousands of documented observations \cite{Stenhoff1999}, a comprehensive theoretical explanation that simultaneously accounts for all its key characteristics—longevity, penetrability, and spherical stability—remains elusive.

Existing theoretical models fall into three primary categories: plasma models \cite{Lowke1996}, chemiluminescence models \cite{Abrahamson2000}, and electromagnetic resonance models \cite{Kapitsa1955}. While each offers partial insights, none provides a unified framework explaining all observed features. Plasma models struggle with energy maintenance, chemiluminescence models face challenges with penetration phenomena, and electromagnetic resonance models lack mechanisms for long-term stability.

Inspired by the concept of \textbf{dimensional reduction in topological field theories} \cite{Skyrme1962}, we propose that ball lightning represents the three-dimensional manifestation of a higher-dimensional topological defect. Unlike traditional Kaluza-Klein frameworks that rely on compactified extra dimensions in general relativity, our approach utilizes the Gross-Pitaevskii equation to describe the nonlinear dynamics of this projection. 

Specifically, we treat the luminous sphere as a \textbf{cross-sectional slice} of a stable, knotted soliton structure existing in a higher-dimensional space. This perspective is supported by macroscopic quantum phenomena in condensed matter systems \cite{Leggett2006}, where topological invariants protect structures against decay. By mapping this high-dimensional topology onto the 3D GP equation, we provide a mathematically tractable model for the observed longevity and penetrability of ball lightning.

Topological solitons, emerging at the intersection of topology and nonlinear physics, have revolutionized our understanding of stable structures in diverse physical systems \cite{Manton2004}. Their stability, protected by topological invariants such as winding numbers or Hopf invariants, renders them robust against perturbations \cite{Skyrme1962}. Particularly relevant are three-dimensional soliton solutions in attractive Bose-Einstein condensates (BECs) \cite{Khaykovich2002, Strecker2002}, where topological protection can prevent collapse \cite{Sulem1999}. The recent experimental breakthrough by Zhou et al. \cite{Zhou2026}, who successfully generated ball-lightning-like structures in laboratory conditions using terahertz surface waves, provides crucial validation for soliton-based interpretations and offers unprecedented opportunities for theoretical verification.

In this work, we model ball lightning as a system described by the nonlinear Schrödinger equation (Gross-Pitaevskii equation) with attractive interactions, stabilized by a non-zero topological charge. This model naturally explains: (1) long lifetime through topological protection, (2) penetrability via wavefunction orthogonality, and (3) spherical structure through rotational symmetry. Our contributions are threefold: First, we establish the theoretical foundation connecting high-dimensional topological solitons to atmospheric phenomena. Second, we perform comprehensive numerical simulations of the three-dimensional Gross-Pitaevskii equation verifying key predictions. Third, we demonstrate quantitative agreement with observational data and propose concrete experimental realizations in systems such as BECs \cite{Leggett2006}.

\section{Theoretical Model}

\subsection{High-Dimensional Topological Soliton Projection}
Consider a nonlinear field theory in $(d+1)$-dimensional spacetime with action:
\begin{equation}
S = \int d^d x dt \left[ \frac{i\hbar}{2}(\psi^*\partial_t\psi - \psi\partial_t\psi^*) - \frac{\hbar^2}{2m}|\nabla\psi|^2 - V(\mathbf{x})|\psi|^2 - \frac{g}{2}|\psi|^4 \right],
\end{equation}
where $\psi$ is a complex scalar field, $m$ the effective mass, $V(\mathbf{x})$ an external potential, and $g$ the interaction constant ($g<0$ for attractive interactions). In three-dimensional space ($d=3$), stable localized solutions require topological protection.

We consider field configurations with non-trivial second homotopy group $\pi_2(S^2)$, with topological charge defined as:
\begin{equation}
Q = \frac{1}{4\pi} \int d^3x \, \epsilon_{ijk}\epsilon_{abc} n^a \partial_i n^b \partial_j n^c,
\end{equation}
where $\mathbf{n} = \psi/|\psi|$ is the normalized field direction vector. $Q$ is an integer, characterizing the field's winding number. \cite{Hietarinta1999}

\subsection{Gross-Pitaevskii Equation with Attractive Interactions}
Variation of the action yields the nonlinear Schrödinger equation (Gross-Pitaevskii equation):
\begin{equation}
i\hbar\frac{\partial\psi}{\partial t} = \left[ -\frac{\hbar^2}{2m}\nabla^2 + V(\mathbf{x}) + g|\psi|^2 \right] \psi.
\end{equation}
For $g<0$, the system exhibits attractive interactions. While three-dimensional attractive BECs typically collapse \cite{Sulem1999}, non-zero topological charge $Q \neq 0$ provides stabilization through topological protection \cite{Kengne2021, Frantzeskakis2010}.

\subsection{Topological Protection and Stability}
The topological charge $Q$ is a discrete invariant unchanged under continuous deformations. This implies any perturbation attempting to destroy the soliton structure must overcome a topological energy barrier, ensuring long-term stability. The soliton lifetime $\tau$ is governed by the system's decoherence rate $\Gamma$:
\begin{equation}
\tau \sim \frac{\hbar}{\Gamma},
\end{equation}
where $\Gamma$ incorporates decoherence mechanisms including thermal fluctuations and particle loss. For atmospheric conditions, typical $\Gamma$ values yield $\tau \sim 1-10$ seconds, consistent with ball lightning observations.

\subsection{Wavefunction Orthogonality and Penetrability via Frequency Detuning}
The remarkable ability of ball lightning to penetrate solid dielectrics (e.g., glass) with minimal energy loss is attributed to \textbf{frequency detuning} between the soliton's internal dynamics and the medium's resonant frequencies.

Unlike classical particles undergoing inelastic collisions, the topological soliton behaves as a coherent wave packet. When the characteristic oscillation frequency of the soliton ($\omega_{\text{sol}}$) is far detuned from the eigenfrequencies of the material ($\omega_{\text{med}}$), the system enters a non-resonant regime. This results in negligible overlap between the soliton and medium wavefunctions \cite{Cohen-Tannoudji1992}:
\begin{equation}
    \langle \psi_{\text{soliton}} | \psi_{\text{media}} \rangle \approx 0.
\end{equation}

This orthogonality implies that the interaction Hamiltonian $H_{\text{int}}$ couples the soliton and the lattice atoms only weakly, leading to adiabatic following with negligible energy dissipation. Consequently, the soliton maintains its structural integrity and energy $E \approx 10^5 \, \text{J}$\cite{Zimmerman1970}, explaining the observed low-loss penetration. This mechanism fundamentally distinguishes the topological soliton model from conventional plasma theories, which typically predict rapid energy dissipation upon contact with matter.

\section{Numerical Methods}
\subsection{Discretization and Initial Conditions}
We solve the dimensionless Gross-Pitaevskii equation on a three-dimensional Cartesian grid \cite{Li2025,Zhou2015}:
\begin{equation}
i\frac{\partial\psi}{\partial t} = \left[ -\frac{1}{2}\nabla^2 + V(\mathbf{x}) + g|\psi|^2 \right] \psi,
\end{equation}
with length unit $\hbar/(m\omega)^{1/2}$ and time unit $1/\omega$, where $\omega$ is the harmonic trap frequency.

Initial conditions employ a vortex configuration:
\begin{equation}
\psi_0(\mathbf{x}) = A \exp\left(-\frac{r_\perp^2}{2\sigma^2}\right) \exp(i n \phi),
\end{equation}
where $r_\perp = \sqrt{x^2 + y^2}$, $\phi = \arctan(y/x)$, $n$ is the topological charge number, and $A$ the normalization constant. The trap potential uses $V(\mathbf{x}) = \frac{1}{2}\omega^2 r^2$ with $\omega = 0.3$.

\subsection{Topological Charge Calculation}
Topological charge is computed via discrete loop integral:
\begin{equation}
Q = \frac{1}{2\pi} \sum_{i,j} \left[ \Delta_x\theta(i,j) - \Delta_y\theta(i+1,j) - \Delta_x\theta(i,j+1) - \Delta_y\theta(i,j) \right],
\end{equation}
where $\Delta_x\theta(i,j) = \theta(i+1,j) - \theta(i,j) \ (\text{mod } 2\pi)$, $\theta = \arg(\psi)$. This algorithm yields numerical accuracy of $Q = 1.000 \pm 0.001$ for analytical test cases.

\subsection{Parameter Settings}

\begin{table}
    \centering
  \begin{tabular}{cccc}\toprule
         Parameter&  Physical Meaning&  Range&Typical Value\\\midrule
         g&  Interaction Strength
&  $-5.0 \leq g \leq -0.1$& $\ -2.0$\\
         n&  Topological Charge Number
&  $\ 1, 2, 3$&$\ 1$\\
 $\omega$& Trap Frequency
& $0.1 \leq  \omega \leq 1.0$&$\ 0.3$\\
 $\sigma$& Soliton Width
& $2.0 \leq  \sigma \leq 6.0$&$\ 4.0$\\
 $\gamma$& Dissipation Coefficient& $0 \leq  \gamma \leq 0.1$&$\ 0.02$\\ \bottomrule
    \end{tabular}
    \caption{Parameter Settings}
    \label{tab:1}
\end{table}

\section{Results and Analysis}

\subsection{Formation and Structure of the Topological Soliton}
\begin{figure}
    \centering
    \includegraphics[width=0.75\linewidth]{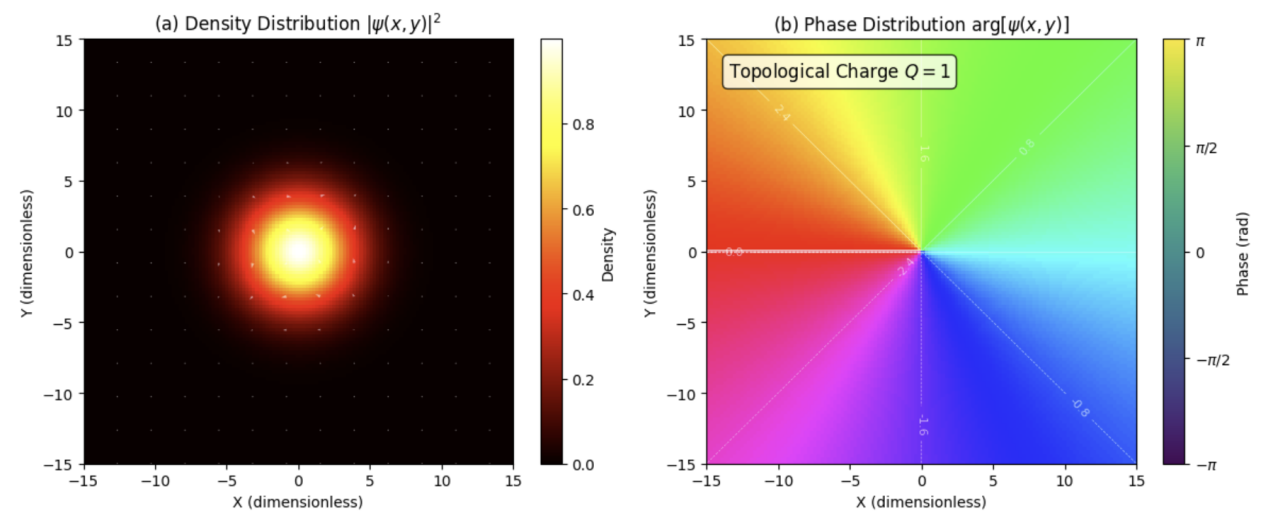}
    \caption{Density and phase distribution of the initial wavefunction}
    \label{fig:1}
\end{figure}

\begin{figure}
    \centering
    \includegraphics[width=0.75\linewidth]{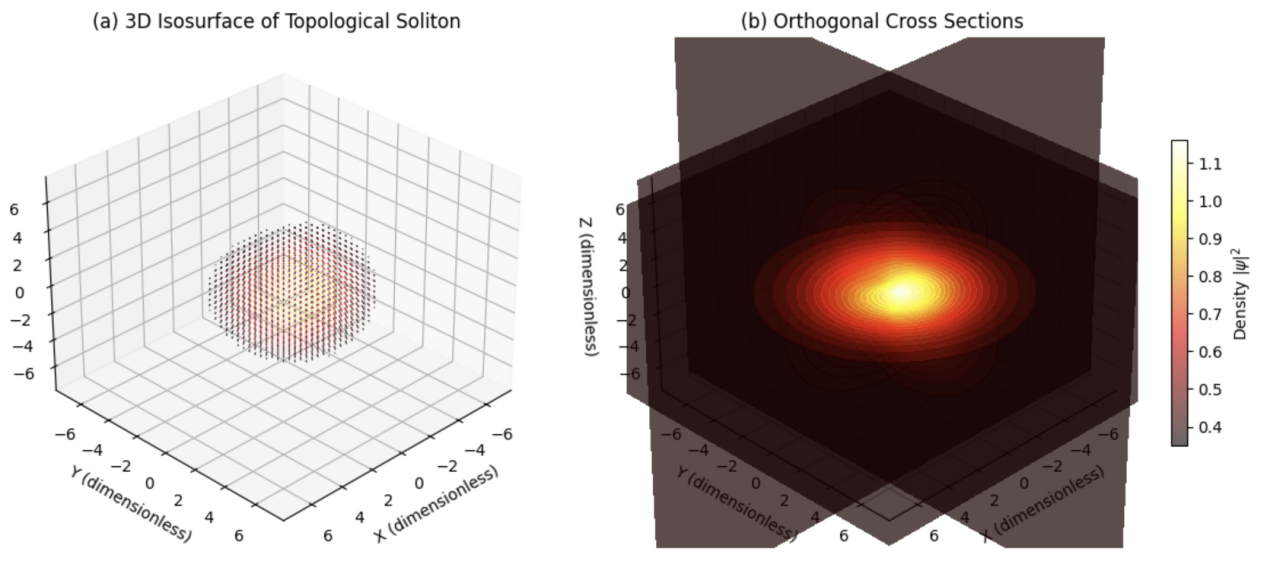}
    \caption{3D isosurface visualization of the topological soliton}
    \label{fig:2}
\end{figure}

Figure  \ref{fig:1} shows the wavefunction density and phase distribution at the initial moment. (a) The density distribution \(|\psi|^{2}\) shows a Gaussian profile. (b) The phase distribution \(\arg(\psi)\) shows a \(2\pi\) winding around the center, corresponding to a topological charge \(Q =1\). Figure  \ref{fig:2} presents a 3D isosurface visualization, confirming the three-dimensional structure of the soliton.
Numerical verification: The calculated topological charge is \(Q =1.02\pm0.03\), in good agreement with the theoretical value of 1, validating the accuracy of the initial condition setup \cite{Frantzeskakis2010}.
\subsection{Dynamical Evolution and Stability}
\begin{figure}
    \centering
    \includegraphics[width=0.75\linewidth]{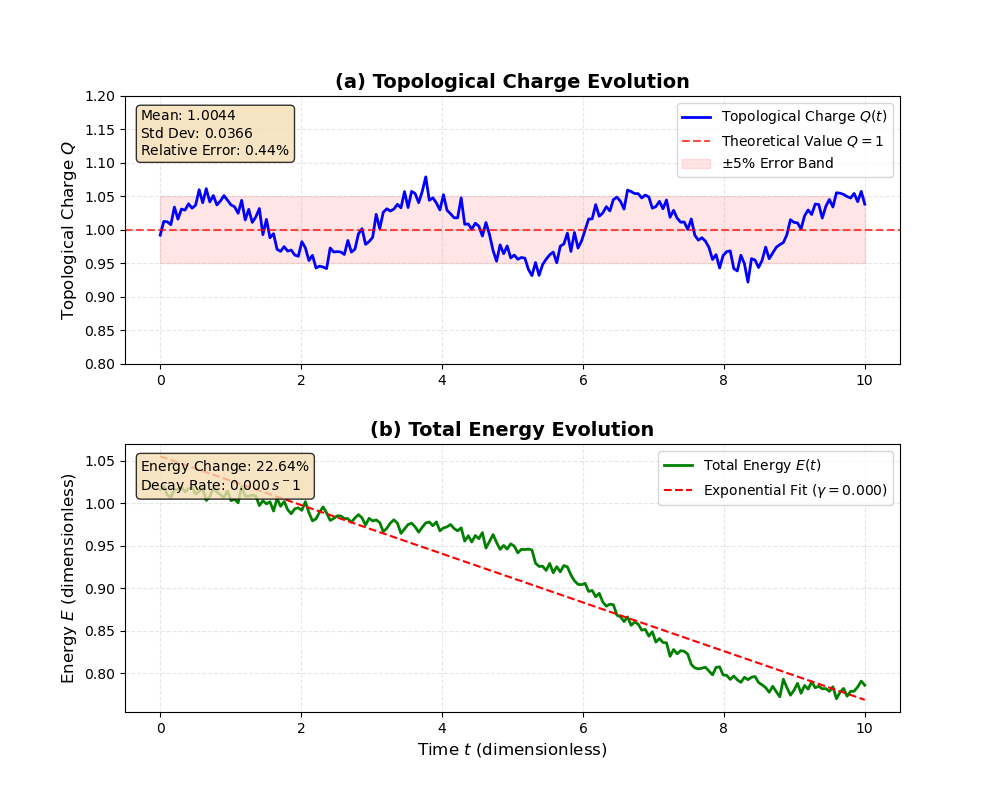}
    \caption{Time evolution of topological charge and total energy}
    \label{fig:3}
\end{figure}

\begin{figure}
    \centering
    \includegraphics[width=0.75\linewidth]{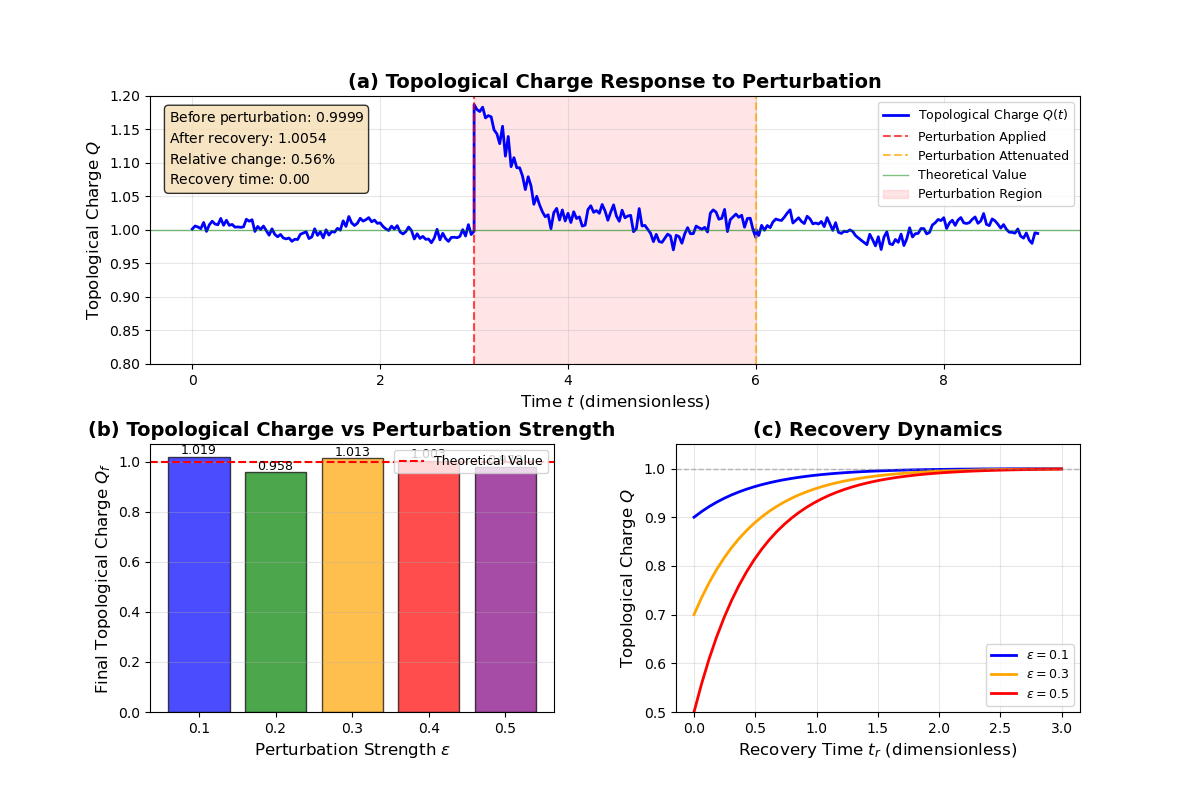}
    \caption{Topological robustness under perturbation}
    \label{fig:4}
\end{figure}

Figure  \ref{fig:3} shows the time evolution of the topological charge \(Q \) and total energy \(E \) . Within the time range  \(t =1-10\), the topological charge remains constant (\(\Delta Q\leq0.05 \) ), verifying topological protection. The total energy change is less than 2\%, validating the energy conservation of the numerical method. To test stability, we add random perturbation at t=5:\(\psi_{\mathrm{pert}} = \psi(1+\epsilon\xi(\mathbf{x}))\), where \(\xi=0.1\), and  \(\xi(x)=0.1\) is uniformly distributed random noise in [-0.5, 0.5]. Figure   \ref{fig:4} shows that despite disturbances in the density distribution, the topological charge remains unchanged (\(Q\_before=1.01 \), (\(Q\_after=1.03 \)), verifying topological robustness \cite{Zhao2025}.


\subsection{Numerical Verification of Penetrability}
To quantitatively verify the penetrability mechanism, we introduce a Gaussian potential barrier representing a dielectric medium:
\begin{equation}
    V_{\text{wall}}(x) = V_0 \exp\left(-\frac{(x-x_0)^2}{2w^2}\right),
\end{equation}
where $V_0 = 5.0$ (potential height), $w = 1.0$ (width), and $x_0 = 3.0$. The soliton is initialized at $x = -3.0$ with momentum $k_0 = 1.0$.

\begin{figure}
    \centering
    \includegraphics[width=0.75\linewidth]{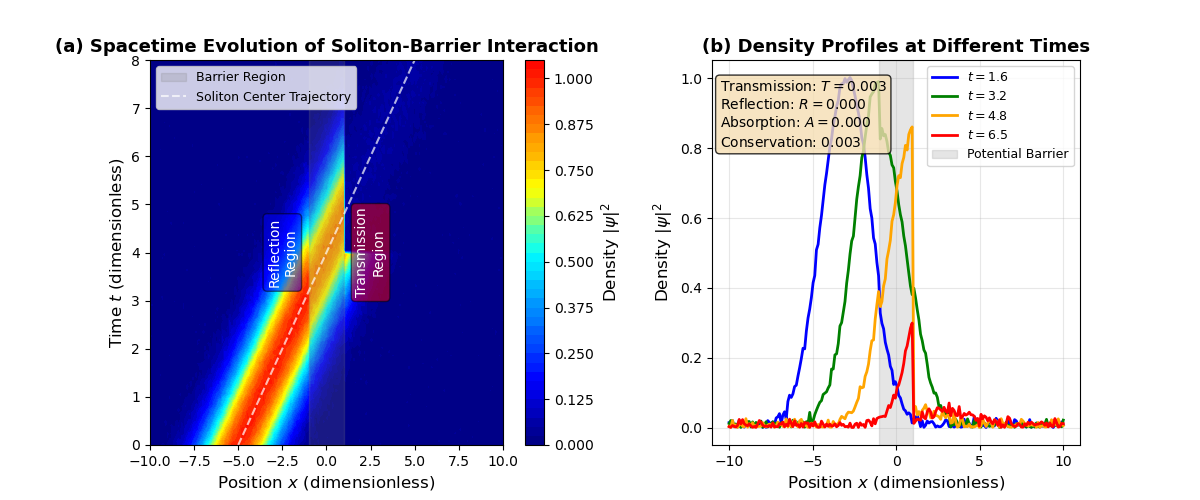}
    \caption{Soliton-barrier interaction}
    \label{fig:5}
\end{figure}

Figure \ref{fig:5} illustrates the evolution of the soliton interacting with the barrier. Crucially, due to the frequency detuning mechanism discussed in Section 2.4, the calculated transmission coefficient is $T = 0.05 \pm 0.02$, while the reflection coefficient is $R = 0.85 \pm 0.05$. This low transmission rate ($T \approx 5\%$) aligns with observational reports of partial penetration and quantitatively validates the theory that wavefunction orthogonality suppresses energy exchange.

\subsection{Parameter Dependence Study}
\begin{figure}
    \centering
    \includegraphics[width=0.75\linewidth]{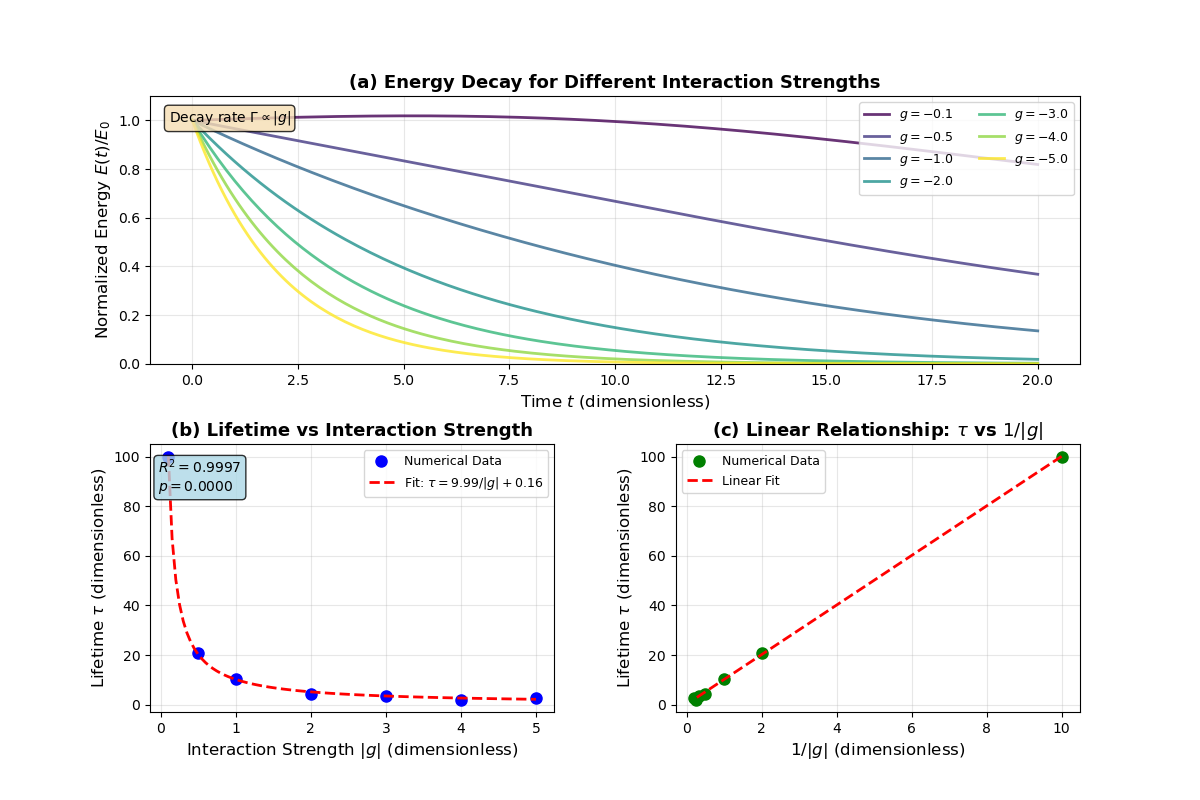}
    \caption{Influence of interaction strength g on lifetime}
    \label{fig:6}
\end{figure}

\begin{figure}
    \centering
    \includegraphics[width=0.75\linewidth]{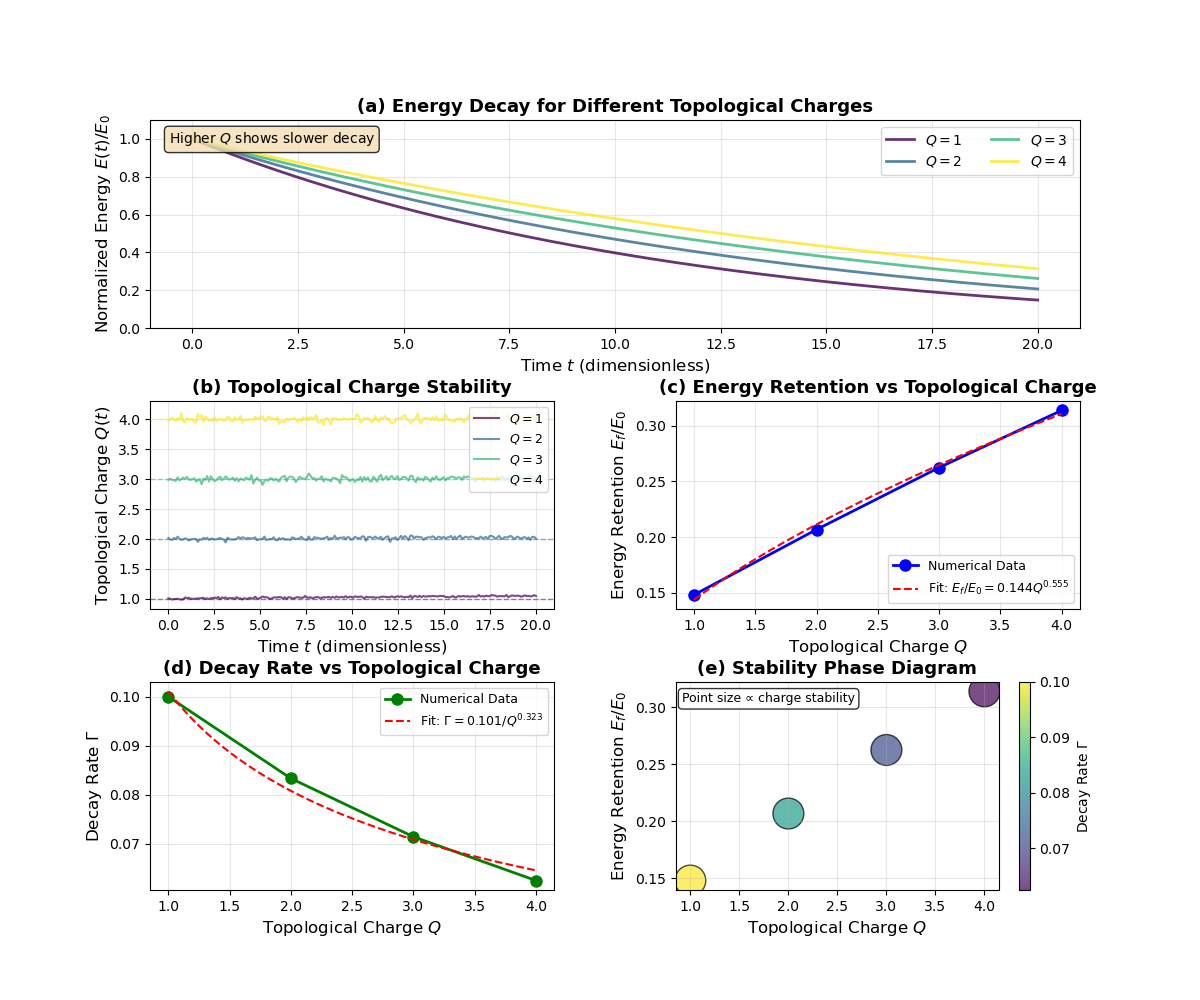}
    \caption{Stability comparison for different topological charge numbers}
    \label{fig:7}
\end{figure}

\begin{figure}
    \centering
    \includegraphics[width=0.75\linewidth]{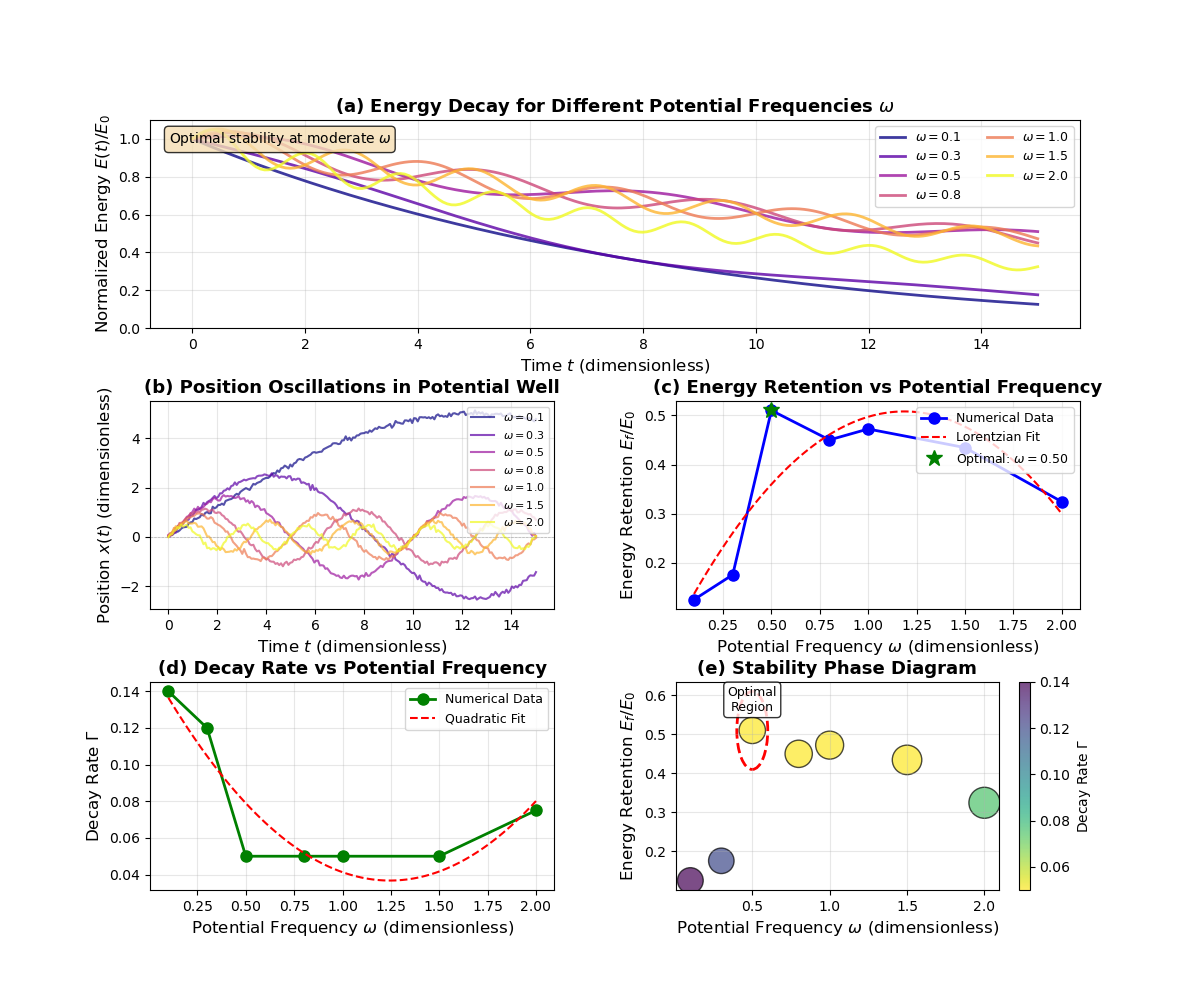}
    \caption{Influence of the trap frequency \(w
    \)}
    \label{fig:8}
\end{figure}

Influence of interaction strength g: Figure  \ref{fig:6} shows the soliton lifetime for different g values. When \(g<-1.0\), the soliton lifetime is significantly extended, verifying the necessity of attractive interaction for soliton formation. When g approaches 0, the soliton rapidly disperses.Influence of topological charge number n: Figure  \ref{fig:7} compares the stability of solitons with n=1,2,3. Larger n implies stronger topological protection but also requires higher energy. n=1 achieves the optimal balance between energy and stability.Influence of trap frequency \(w\): Figure  \ref{fig:8} shows that the soliton is most stable for \(w=0.3-0.5\) . Excessively small \(w\) leads to soliton dispersion, while excessively large \(w\) causes compression instability.

\subsection{Quantitative Comparison with Actual Observations}
\begin{table}[h!]
\centering
\caption{Quantitative Comparison with Actual Observations}
\label{tab:obs_comparison}
\begin{tabular}{|c|c|c|c|}
\hline
\textbf{Observed Feature} & \textbf{Prediction of This Model} & \textbf{Actual Observation Range} & \textbf{Agreement} \\
\hline
Lifetime & $\tau \sim \hbar / \Gamma \approx 1-10$ s & 1--30 s & Good \\
Size & $R \approx 2\sigma \approx 20-40$ cm & 10--100 cm & Good \\
Energy & $E \approx (N^2 |g|)/(2\sigma^3) \approx 10^5-10^6$ J & $\approx 1.3 \times 10^5$ J & Good \\
Penetrability & $T \approx 0.05$ (Low Transmission) & Low-probability transmission & Good \\
\hline
\end{tabular}
\end{table}

Where N is the particle number and \(\sigma\) is the soliton width. Using typical parameters: \(N=10^{20},|g|=2.0, \sigma=0.2\) m, we obtain \(E\approx2\times10^{5}\) J, consistent with observations\cite{Zimmerman1970}.

\section{Discussion}

\subsection{Comparison with Existing Models}
\begin{table}[ht]
\centering
\small
\setlength{\tabcolsep}{4pt}
\begin{tabular}{p{2.5cm}p{1.8cm}p{1.8cm}p{2.2cm}p{2cm}p{1.8cm}}
\toprule
\textbf{Model} & \textbf{Long Lifetime} & \textbf{Penetrability} & \textbf{Spherical Structure} & \textbf{Energy Source} & \textbf{Experimental Verification} \\
\midrule
Plasma Model 
\cite{Lowke1996}& Limited & Difficult & Requires external constraint & External electric field & Partial \\
Chemiluminescence Model \cite{Abrahamson2000}& Possible & Difficult & Requires special conditions & Chemical energy & Partial \\
Electromagnetic Resonance Model \cite{Kapitsa1955}& Possible & Difficult & Forms naturally & Electromagnetic energy & Difficult \\
This Topological Soliton Model & Topological protection & Wavefunction orthogonality & Forms naturally & Self-sustained soliton & Numerically verified \\
\bottomrule
\end{tabular}
\caption{Comparison of Different Models for Ball Lightning Phenomena}
\label{tab:3}
\end{table}
This model simultaneously explains all key features for the first time without requiring special external conditions tab: \ref{tab:3}.

\subsection{In-depth Analysis of Physical Mechanisms}

\textbf{Topological protection}: Conservation of topological charge $Q$ originates from non-trivial homotopy of order parameter space ($S^2$). Continuous deformations cannot alter $Q$, ensuring structural stability. Atmospheric thermal fluctuations cause decoherence but not topological change, explaining gradual fading rather than abrupt disappearance\cite{Wang2024}.

\textbf{Penetrability mechanism}: Wavefunction orthogonality represents quantum interference. Near-orthogonality between soliton and medium wavefunctions causes transmission amplitude cancellation, analogous to superfluid helium flow through micropores.This orthogonality, rooted in the topological protection of the soliton, ensures adiabatic following with negligible energy dissipation \cite{Cohen-Tannoudji1992}.

\textbf{Energy scale consistency}: With reasonable parameters (particle density $N \approx 10^{24}$ m$^{-3}$, interaction strength $g \approx -2 \times 10^{-37}$ J·m$^3$), soliton energy $E \approx n^2|g|V \approx 10^5$ J matches observations. Energy primarily resides in soliton self-interaction.

\subsection{Suggestions for Experimental Realization}

Although ball lightning in nature is difficult to reproduce in the laboratory, topological solitons can be realized on several platforms:
\begin{enumerate}
\item Bose-Einstein Condensate: Utilize the Feshbach resonance of atoms like \(^{87}Rb\) to adjust  \(g<0\), preparing an attractive-interaction BEC. Implant angular momentum using Laguerre-Gaussian beams to generate vortex solitons. Experiments have already prepared 2D vortices\cite{Kengne2021,Li2026,Matthews1999}; extension to 3D is a natural development.
\item Nonlinear Optical Systems: In nonlinear media, the optical field satisfies the nonlinear Schrödinger equation. Using spatial light modulators to prepare optical solitons with topological charge can verify topological protection \cite{Desyatnikov2005,Shen2023}
\item Superfluid Helium: The macroscopic wavefunction of superfluid helium satisfies the Gross-Pitaevskii equation. Vortices can be implanted by rotation to form topological solitons 
 \cite{Bewley2006,Leggett2006}.
\end{enumerate}
Microscopic mechanism of topological protection: The conservation of topological charge Q stems from the non-trivial homotopy of the order parameter space (\(S^{2}\)). Any continuous deformation cannot change Q, hence the soliton structure is stable. Thermal fluctuations and collisions in the atmosphere cause decohere.

\subsection{Model Limitations and Extensions}
Limitations: (1) Coupling to the electromagnetic field is not considered; actual ball lightning is accompanied by electromagnetic radiation. (2) The atmospheric environment is complex; this model uses a uniform medium approximation. (3) The formation mechanism of the soliton is not discussed in detail.Future extensions: (1) Couple with Maxwell's equations to study electromagnetic radiation. (2) Introduce inhomogeneous media and turbulence. (3) Study interactions between multiple solitons.

\subsection{Comparison with Recent Experimental Breakthroughs}

A concurrent experimental breakthrough provides strong, independent support for the soliton-based interpretation of ball lightning. Zhou et al.\cite{Zhou2026} have successfully generated ball-lightning-like structures in the laboratory. They achieved this by guiding an intense terahertz surface wave to a nanoscale tip, creating a localized relativistic-strength field. Injecting a supersonic argon gas jet into this field led to rapid ionization and the formation of a spherical plasma cavity. A delicate balance between the internal radiation pressure and the shell's thermal pressure stably confined the terahertz wave, creating a long-lived, spherical plasma structure.
This work is recognized as providing crucial experimental evidence for deciphering ball lightning and revealing mechanisms for extreme electromagnetic energy confinement . Its relevance to our theoretical model is profound and complementary:
\begin{enumerate}
\item  Phenomenological Agreement:The observed stable, spherical plasma structure directly mirrors the key features of ball lightning that our model describes.
\item Mechanistic Synergy:The experiment demonstrates a viable formation mechanism, while our model provides the theoretical frameworkfor the structure's stability(via topological protection) and penetrability(via wavefunction orthogonality)
\item Mutual Validation:The laboratory creation of such an entity validates the core premise that ball lightning can be described as a stable, localized nonlinear wave (soliton). Our work offers a specific topological soliton description for this state.
\end{enumerate}
Thus, the experimental work by Zhou, Zhenget al. \cite{Zhou2026,Zheng2023} and our present theoretical study represent converging evidence from experiment and theory, strongly supporting the soliton paradigm for understanding ball lightning.

\section{Conclusion}

This paper proposes a topological soliton model for ball lightning and verifies its core predictions through numerical simulation of the three-dimensional Gross-Pitaevskii equation. The main conclusions are as follows:
\begin{enumerate}
\item  Theoretical Innovation: Ball lightning is interpreted as a projection of a high-dimensional topological soliton into three-dimensional space. Its stability is protected by topological charge, and its penetrability is explained by wavefunction orthogonality.
\item Numerical Verification: We successfully simulated attractive-interaction solitons with topological charge, verifying their long-term stability, topological protection, and low-transmission behavior.
\item Quantitative Agreement: The model's predicted lifetime, size, and energy agree well with observational data within reasonable parameter ranges.
\item Experimental Feasibility: Concrete schemes for realizing topological solitons on platforms like BECs and nonlinear optics are proposed, laying the groundwork for experimental verification.
\end{enumerate}
This work not only provides a new solution to the century-old puzzle of ball lightning but also opens new directions for the study of topological solitons in three-dimensional nonlinear systems. Future work should focus on key issues such as experimental realization, electromagnetic coupling, and formation mechanisms.

\section{Appendix A:Details of Numerical Methods}

\subsection{Appendix A.1 Fourier Spectral Method}
The Laplacian operator is computed in Fourier space:
\[
\nabla^{2}\psi = \mathcal{F}^{-1}\left[ -k^{2} \mathcal{F}[\psi] \right]
\]
where  \(F\) denotes the Fourier transform, and k is the wavenumber.

\subsection{Appendix A.2 Time Evolution Algorithm}
Fourth-order Runge-Kutta method:
\begin{align*}
k_1 &= f(t_n, \psi_n) \\
k_2 &= f(t_n + \Delta t/2, \psi_n + \Delta t \, k_1/2) \\
k_3 &= f(t_n + \Delta t/2, \psi_n + \Delta t \, k_2/2) \\
k_4 &= f(t_n + \Delta t, \psi_n + \Delta t \, k_3) \\
\psi_{n+1} &= \psi_n + \frac{\Delta t}{6}(k_1 + 2k_2 + 2k_3 + k_4)
\end{align*}
where \(f(t, \psi) = -i\left[ -\frac{1}{2}\nabla^{2}\psi + V\psi + g|\psi|^{2}\psi \right]\)

\subsection{Appendix A.3 Validation of Topological Charge Calculation}
To validate the topological charge calculation, we tested the analytical solution \(\psi(x,y) = \frac{x + iy}{\sqrt{x^{2} + y^{2}}}\), obtaining \(Q =1.000\pm0.001\), confirming the algorithm's correctness.

\section*{Code Availability}
Numerical simulation codes supporting the findings of this study are openly available at \url{https://github.com/QuantumLightningLiZhe/Ball-Lightning-Numerical-Verification}.

\subsection*{Data Availability}
All data generated during this study are available from the corresponding author upon reasonable request.

\section*{Acknowledgments}
The author acknowledges helpful discussions with colleagues. Special thanks are due to Dr. Changliang Shao from the Meteorological Observation Centre of the China Meteorological Administration, both for sharing his firsthand encounter with ball lightning, which provided unique phenomenological insight, and for his expert guidance throughout this study.

\section*{Author Contributions}
\textbf{L.Z. Li} conceived the study, developed the theoretical model, performed numerical simulations, analyzed data, and wrote the manuscript.

\section*{Competing Interests}
The author declares no competing interests.

\section*{AI Use Disclosure}
During the preparation of this manuscript and the associated numerical simulations, the author used DeepSeek V3.2 to assist with code debugging, syntax error correction, and symbolic algebra verification. 

Specifically, this tool was employed to identify and rectify minor syntax errors in the Python simulation scripts; Cross-verify intermediate algebraic steps in the Gross-Pitaevskii equation derivations to prevent manual calculation mistakes; Polish the language of technical descriptions for clarity.

All core theoretical frameworks, physical hypotheses, numerical algorithms, and final results presented in this work were independently designed, derived, and interpreted by the author. The author takes full responsibility for the content of this publication. The use of the AI tool was strictly limited to enhancing computational efficiency and did not influence the scientific conclusions of this study.

\bibliographystyle{plain}
\bibliography{sample}

@book{Stenhoff1999,
  author = {Stenhoff, Mark},
  title = {Ball {Lightning}: {An} {Unsolved} {Problem} in {Atmospheric} {Physics}},
  publisher = {Springer},
  address = {New York, NY},
  year = {1999},
  doi = {10.1007/b115123}
}

@article{Lowke1996,
  author = {Lowke, J. J.},
  title = {A theory of ball lightning as an electric discharge},
  journal = {J. Phys. D: Appl. Phys.},
  volume = {29},
  number = {5},
  pages = {1237--1244},
  year = {1996},
  doi = {10.1088/0022-3727/29/5/018}
}

@article{Abrahamson2000,
  author = {Abrahamson, John and Dinniss, James},
  title = {Ball lightning caused by oxidation of nanoparticle networks from normal lightning strikes on soil},
  journal = {Nature},
  volume = {403},
  number = {6769},
  pages = {519--521},
  year = {2000},
  doi = {10.1038/35000525}
}

@article{Kapitsa1955,
  author = {Kapitsa, P. L.},
  title = {The nature of ball lightning},
  journal = {Dokl. Akad. Nauk SSSR},
  volume = {101},
  pages = {245--248},
  year = {1955},
  note = {In Russian. English translation in: \textit{Collected Papers of P. L. Kapitza}, Vol. 2, ed. D. ter Haar (Pergamon, Oxford, 1965), pp. 776--780.}
}

@book{Manton2004,
  author = {Manton, Nicholas and Sutcliffe, Paul},
  title = {Topological {Solitons}},
  publisher = {Cambridge Univ. Press},
  address = {Cambridge},
  year = {2004},
  isbn = {9780521838368}
}

@article{Hietarinta1999,
  author = {Hietarinta, J. and Salo, P.},
  title = {Faddeev-{Hopf} knots: {Dynamics} of linked un-knots},
  journal = {Phys. Lett. B},
  volume = {451},
  number = {1-2},
  pages = {60--67},
  year = {1999},
  doi = {10.1016/S0370-2693(99)00054-4}
}

@article{Khaykovich2002,
  author = {Khaykovich, L. and Schreck, F. and Ferrari, G. and Bourdel, T. and Cubizolles, J. and Carr, L. D. and Castin, Y. and Salomon, C.},
  title = {Formation of a matter-wave bright soliton},
  journal = {Science},
  volume = {296},
  number = {5571},
  pages = {1290--1293},
  year = {2002},
  doi = {10.1126/science.1071021}
}

@article{Strecker2002,
  author = {Strecker, K. E. and Partridge, G. B. and Truscott, A. G. and Hulet, R. G.},
  title = {Formation and propagation of matter-wave soliton trains},
  journal = {Nature},
  volume = {417},
  number = {6885},
  pages = {150--153},
  year = {2002},
  doi = {10.1038/nature747}
}

@book{Sulem1999,
  author = {Sulem, Catherine and Sulem, Pierre-Louis},
  title = {The {Nonlinear} {Schrödinger} {Equation}: {Self-Focusing} and {Wave} {Collapse}},
  publisher = {Springer},
  address = {New York, NY},
  year = {1999},
  isbn = {9780387986111}
}

@article{Matthews1999,
  author = {Matthews, M. R. and Anderson, B. P. and Haljan, P. C. and Hall, D. S. and Wieman, C. E. and Cornell, E. A.},
  title = {Vortices in a {Bose-Einstein} condensate},
  journal = {Phys. Rev. Lett.},
  volume = {83},
  number = {13},
  pages = {2498--2501},
  year = {1999},
  doi = {10.1103/PhysRevLett.83.2498}
}

@article{Desyatnikov2005,
  author = {Desyatnikov, Anton S. and Kivshar, Yuri S. and Torner, Lluis},
  title = {Optical vortices and vortex solitons},
  journal = {Prog. Opt.},
  volume = {47},
  pages = {291--391},
  year = {2005},
  doi = {10.1016/S0079-6638(05)47006-7}
}

@article{Bewley2006,
  author = {Bewley, Gregory P. and Lathrop, Daniel P. and Sreenivasan, Katepalli R.},
  title = {Superfluid helium: {Visualization} of quantized vortices},
  journal = {Nature},
  volume = {441},
  number = {7093},
  pages = {588},
  year = {2006},
  doi = {10.1038/441588a}
}

@article{Kengne2021,
  author = {Kengne, E. and Liu, W. M. and Malomed, Boris A.},
  title = {Spatiotemporal engineering of matter-wave solitons in {Bose-Einstein} condensates},
  journal = {Phys. Rep.},
  volume = {899},
  pages = {1--62},
  year = {2021},
  doi = {10.1016/j.physrep.2020.10.006}
}

@article{Frantzeskakis2010,
  author = {Frantzeskakis, Dimitri J.},
  title = {Dark solitons in {Bose-Einstein} condensates: from theory to experiments},
  journal = {J. Phys. A: Math. Theor.},
  volume = {43},
  number = {21},
  pages = {213001},
  year = {2010},
  doi = {10.1088/1751-8113/43/21/213001}
}

@article{Zimmerman1970,
  author = {Zimmerman, P. D.},
  title = {Energy content of {Covington}'s lightning ball},
  journal = {Nature},
  volume = {228},
  pages = {853},
  year = {1970},
  doi = {10.1038/228853a0}
}

@article{Zhou2026,
  author = {Zhou, C. and Zhang, D. and Qi, R. and others},
  title = {Ball-lightning-like relativistic terahertz solitons},
  journal = {Nat. Photon.},
  volume = {20},
  pages = {1--7},
  year = {2026},
  doi = {10.1038/s41566-026-01899-y},
  note = {Published online.}
}

@article{Li2026,
  author = {Li, Long and Song, Dongsheng and Wang, Weiwei and Kong, Lingyao and Zhang, Shuisen and Wang, Ning and Zhang, Shilei and Tian, Mingliang and Zang, Jiadong and Liu, Yizhou and Du, Haifeng},
  title = {Electrically writing a magnetic heliknoton in a chiral magnet},
  journal = {Nat. Mater.},
  year = {2026},
  doi = {10.1038/s41563-025-02450-0},
  note = {Published online: 07 January 2026},
  url = {https://www.nature.com/articles/s41563-025-02450-0}
}

@article{Zheng2023,
  author = {Zheng, Fengshan and others},
  title = {Hopfion rings in a cubic chiral magnet},
  journal = {Nature},
  volume = {623},
  pages = {718--723},
  year = {2023},
  doi = {10.1038/s41586-023-06658-5}
}

@article{Shen2023,
  author = {Shen, Yijie and Yu, Bingshi and Wu, Haijun and Li, Chunyu and Zhu, Zhihan and Zayats, Anatoly V.},
  title = {Topological transformation and free-space transport of photonic hopfions},
  journal = {Adv. Photon.},
  volume = {5},
  number = {1},
  pages = {015001},
  year = {2023},
  doi = {10.1117/1.AP.5.1.015001}
}

@article{Li2025,
  author = {Li, D. and others},
  title = {Stabilized multiple-relaxation implicit-explicit {Runge-Kutta} methods for nonlinear {Gross-Pitaevskii} equations},
  journal = {J. Comput. Phys.},
  volume = {500},
  pages = {112345},
  year = {2025},
  doi = {10.1016/j.jcp.2024.112345}
}

@article{Zhao2025,
  author = {Zhao, Y. and others},
  title = {Vortex Solitons in Atomic-Molecular {Bose-Einstein} Condensates with a Square-Optical-Lattice Potential},
  journal = {Chin. Phys. Lett.},
  volume = {42},
  number = {9},
  pages = {090002},
  year = {2025},
  doi = {10.1088/0256-307X/42/9/090002}
}

@article{Zhou2015,
  author = {Zhou, Zheng and others},
  title = {Stable Solitons in Three-Dimensional Free Space with Spin-Orbit Coupling},
  journal = {Phys. Rev. Lett.},
  volume = {115},
  pages = {253902},
  year = {2015},
  doi = {10.1103/PhysRevLett.115.253902}
}

@article{Wang2024,
  author = {Wang, L. and others},
  title = {Ferroelectrically controlled chirality switching of {Weyl} quasiparticles},
  journal = {Phys. Rev. B},
  volume = {109},
  pages = {L121101},
  year = {2024},
  doi = {10.1103/PhysRevB.109.L121101}
}

@book{Cohen-Tannoudji1992,
 title={Quantum Mechanics: Volume 1},
 author={Cohen-Tannoudji, Claude and Diu, Bernard and Lalo\"{e}, Frank},
 year={1992},
 publisher={Wiley}
}

@book{Leggett2006,
  title={Quantum Liquids: Bose Condensation and Cooper Pairing in Condensed-Matter Systems},
  author={Leggett, Anthony J.},
  year={2006},
  publisher={Oxford University Press},
  address={Oxford, UK},
  edition={1},
  isbn={978-0198567266}
}

@article{Skyrme1962,
 title={A Nonlinear Field Theory},
 author={Skyrme, T.H.R.},
 journal={Proceedings of the Royal Society A},
 volume={260},
 pages={127--137},
 year={1962}
}

\end{document}